Superconductivity Above 40 K Observed in a New Iron Arsenide Oxide $(Fe_2As_2)(Ca_4(Mg,Ti)_3O_y)$

Hiraku Ogino[1,2*], Yasuaki Shimizu[1], Koichi Ushiyama[1,2], Naoto Kawaguchi[1,2], Kohji Kishio[1,2] and Jun-ichi Shimoyama[1,2]

[1]Department of Applied Chemistry, The University of Tokyo, 7-3-1 Hongo, Bunkyo, Tokyo 113-8656, Japan

[2]JST-TRIP, Sanban, Chiyoda, Tokyo 102-0075, Japan

A new layered iron arsenide oxide $(Fe_2As_2)(Ca_4(Mg,Ti)_3O_y)$ was discovered. Its crystal structure is tetragonal with a space group of $I4/mmm$ consisted of the anti-fluorite type FeAs layer and blocking layer of triple perovskite cells and is identical with $(Fe_2As_2)(Sr_4(Sc,Ti)_3O_8)$ discovered in our previous study. The lattice constants of $(Fe_2As_2)(Ca_4(Mg,Ti)_3O_y)$ are $a$ = 3.877 Å and $c$ = 33.37 Å. This compound exhibited bulk superconductivity up to 43 K in susceptibility measurement without intentional carrier doping. A resistivity drop was observed at ~47 K and zero resistance was achieved at 42 K. These values correspond to the second highest $T_c$ among the layered iron-based superconductors after $RE$FeAsO system.

* E-mail address: tuogino@mail.ecc.u-tokyo.ac.jp

Since the discovery of superconductivity in LaFeAs(O,F)[1], a large number of layered compounds containing anti-fluorite type Fe$Pn$ ($Pn$ = pnictogens) layers have been discovered. Among them, a series of pnictide oxides having perovskite-type blocking layers [2-10] are attractive because of their highly two dimensional crystal structure and chemical and structural flexibility at the blocking layers. The latter feature may lead to discoveries of new superconducting materials. In fact, several new compounds have been found in this system, such as $(Fe_2As_2)(Sr_3Sc_2O_5)$[2], $(Fe_2P_2)(Sr_4Sc_2O_6)$[3], $(Fe_2As_2)(Sr_4M_2O_6)$ ($M$ = Sc[4], Cr[4], V[5], MgTi[6]), and $(Fe_2As_2)(Ba_4Sc_2O_6)$[7]. $(Fe_2As_2)(Sr_4V_2O_6)$ and $(Fe_2As_2)(Sr_4(Mg,Ti)_2O_6)$ show superconductivity with $T_{c(onset)}$ of ~ 40 and 42~46 K under ambient pressure and a high pressure of 4 GPa[8], respectively. Such relatively high $T_c$ under high pressure suggested further enhancement of $T_c$ under ambient pressure is possible by improving the local structure of the FeAs layer in this system.

In addition, the compounds with "22438" structure, such as $(Fe_2As_2)(Ca_4(Sc,Ti)_3O_y)$[9], $(Fe_2As_2)(Sr_4(Sc,Ti)_3O_8)$, and $(Fe_2As_2)(Ba_4Sc_3O_{7.5})$[10] were discovered in our recent study. Based on these backgrounds, we have attempted to develop new pnictide oxides having the 22438 structure and discovered a new iron pnictide $(Fe_2As_2)(Ca_4(Mg,Ti)_3O_y)$ in the present study.

Sintered bulk samples with nominal compositions of $(Fe_2As_2)(Ca_4(Mg_{1-x}Ti_x)_3O_{7.5})$ ($0.5 \leq x \leq 0.9$) were synthesized by solid-state reaction starting from FeAs(3N), CaO(3N), MgO(3N), Ti(3N), and $TiO_2$(3N). The starting reagents were moisture sensitive, so that manipulation was carried out in a glove box filled with argon gas. Powder mixtures were pelletized, sealed in evacuated quartz ampoules and heated at 1000-1100°C for 60-100 h followed by slow cooling to room temperature.

Phase identification was carried out by the powder X-ray diffraction (XRD) method using MAC Science MXP-18. The XRD intensity data were collected in the $2\theta$ range of 3-80° at a step of 0.02° using Cu-$K_\alpha$ radiation. Silicon powder was used as an internal standard. Magnetic susceptibility was measured using a superconducting quantum interference device (SQUID) magnetometer (Quantum Design MPMS-XL5s). Electrical resistivity was examined by the AC four-point-probe method under magnetic fields up to 9 T using Quantum Design Physical Properties Measurement System.

Powder XRD pattern of a sample with a nominal composition of $(Fe_2As_2)(Ca_4(Mg_{0.25}Ti_{0.75})_3O_{7.5})$ reacted at 1000ºC for 100 h is shown in Fig. 1 together with the simulation pattern. The 22438 phase formed as a main phase with weak peaks due to impurities, such as $CaTiO_3$ and FeAs. Note that any peaks originated from $CaFe_2As_2$ were not detected. In addition, similar to $(Fe_2As_2)(Sr_4(Mg_{0.5}Ti_{0.5})_2O_6)$[6], any satellite peaks indicating superstructure by the ordering of Mg and Ti were not found in the XRD pattern. It should be noted that the relative peak intensity of impurity phases in the powder XRD patterns were minimum in a sample starting from $x$ = 0.75 and they monotonically increased either by increasing or by decreasing $x$ from 0.75. Moreover, apparent dependences of the lattice constants on $x$ were not confirmed. These suggests that the range of $x$ in the resulting $(Fe_2As_2)(Ca_4(Mg_{1-x}Ti_x)_3O_y)$ phase is quite narrow around 0.75 or slightly less, because a certain amount of $CaTiO_3$ is co-existed. Figure 1(b) shows the crystal structure of the new compound consist of the anti-fluorite type FeAs layer and blocking layer with triple perovskite cells. The space group of the compound is $I4/mmm$ and the lattice constants are $a = 3.877$ Å and $c = 33.37$ Å. These values are slightly shorter than those of $(Fe_2As_2)(Ca_4(Sc_{0.67}Ti_{0.33})_3O_y)$, which has $a = 3.922$ Å and $c = 33.56$ Å, mainly due to higher concentration of $Ti^{4+}$ in the new compound. Although phase purity of the present sample was not enough for further precise analyses, the distance between arsenic and iron-plane, which is called "pnictogen height", is estimated to be approximately 1.4 Å from the $a$-axis length[11]. This value is almost optimum for the high $T_c$ in iron pnictides[12].

In our preliminary study, synthesis of samples with nominal compositions of $(Fe_2As_2)(Ca_4(Mg_{1-x}Ti_x)_3O_8)$ were attempted, because the oxygen content $y$ is 8 when all oxygen sites are fully occupied as shown in Fig. 1(b). These samples were always very low quality containing large amounts of $CaTiO_3$ and other phases as in the case of $(Fe_2As_2)(Ca_4(Sc_{0.67}Ti_{0.33})_3O_8)$[9]. The optimal $y$ in the nominal composition is 7.5 to form $(Fe_2As_2)(Ca_4(Mg_{1-x}Ti_x)_3O_y)$ phase as a major one at the present stage. However, we believe that the oxygen contents in the resulting samples are close to 8, because a part of inner wall of quartz ampoule was found to turn brown due to reduction of $SiO_2$ after the heat-treatment.

Temperature dependences of zero-field-cooled (ZFC) and field-cooled (FC) magnetization of $(Fe_2As_2)(Ca_4(Mg_{0.25}Ti_{0.75})_3O_y)$ are shown in Fig. 2. The sample showed superconducting transition at 43 K. The superconducting volume fraction estimated from the ZFC magnetization at 2 K was much larger than 100%. This very large diamagnetism is due to the porous microstructure having a large amount of closed pores and a demagnetization effect of the sample. The reversible region, where the ZFC and FC magnetization curves overlap each other is quite narrow down to 42 K, suggesting the strong grain coupling compared with other layered iron pnictides having the perovskite-type oxide blocking layers[5,6,9].

Figure 3 shows the temperature dependence of the resistivity for $(Fe_2As_2)(Ca_4(Mg_{0.25}Ti_{0.75})_3O_y)$ measured under various magnetic fields. A resistivity curve up to 300 K under 0 T is shown in the inset. The normal state resistivity exhibited a metallic behavior with a convexity. The superconducting transition was observed at $T_{c(onset)}$ of 47 K and zero resistivity was achieved at 42 K. The transition width ~5 K was narrower than other related compounds having perovskite-type blocking oxide layers[5,6,9], also indicating better grain coupling.

The zero resistivity temperature, $T_{c(\rho=0)}$, was largely and systematically lowered by increasing external fields. Poor flux pinning due to the thick blocking layer, which is believed to enhance the electromagnetic anisotropy, might be an main origin for large broadening at low resistivity region. On the other hand, resistivity curves showed small broadening near $T_{c(onset)}$, suggesting very high upper critical fields as observed in other layered iron arsenides, such as NdFeAs(O,F) and SmFeAs(O,F)[13].

Including a new compound found in the present study, an interesting tendency is revealed when we compare $T_{c(onset)}$ and $a$-axis length of layered iron arsenide oxides with thick blocking layers. $(Fe_2As_2)(Ca_4(Sc,Ti)_3O_y)$, $(Fe_2As_2)(Ca_5(Sc,Ti)_4O_y)$, $(Fe_2As_2)(Ca_6(Sc,Ti)_5O_y)$[9] and $(Fe_2As_2)(Ca_4(Mg,Ti)_3O_y)$ showed $T_{c(onset)}$s at 33, 41, 42, and 47 K, respectively, in the resistivity measurements and their $a$-axis lengths are 3.922, 3.902, 3.884, and 3.877 Å, respectively. It is clear that shortening the $a$-axis length accompanies enhancement of $T_c$. The dependence of $T_c$ on $a$-axis length must be related to the systematic change of local structure in the FeAs layer particularly to the pnictogen height of these compounds.

In summary, a new layered iron pnictide oxide $(Fe_2As_2)(Ca_4(Mg_{1-x}Ti_x)_3O_y)$ has been successfully synthesized by solid-state reaction in quartz ampoule. This phase formed as a main one starting from $x = 0.75$ and $y = 7.5$, respectively, in the nominal composition. Crystal structure of the new compound consists of alternate stacking of anti-fluorite FeAs layer and triple perovskite-type oxide layers. Bulk superconductivity was observed in susceptibility measurement with $T_c$ up to 43 K. $T_{c(onset)}$ and $T_{c(\rho=0)}$ were 47 and 42 K in the resistivity measurement. The relatively high $T_c$ above 40 K of this compound is believed to be attributable to moderately short $a$-axis length, which may results in the nearly optimal "pnictgen height".

**Acknowledgements**
This work was partly supported by a Grant-in-Aid for Young Scientists (B) No. 21750187, 2009, supported by the Ministry of Education, Culture, Sports, Science and Technology, Japan (MEXT) as well as inter-university Cooperative Research Program of the Institute for Materials Research, Tohoku University.

Figure captions

Figure 1. Powder XRD, simulation pattern(a) and crystal structure(b) of $(Fe_2As_2)(Ca_4(Mg_{1-x}Ti_x)_3O_8)$.

Figure 2. Temperature dependence of ZFC and FC magnetization curves under 1 Oe of a sintered bulk sample with a nominal composition of $(Fe_2As_2)(Ca_4(Mg_{0.25}Ti_{0.75})_3O_{7.5})$.

Figure 3. Temperature dependences of resistivity of a sintered bulk sample with a nominal composition of $(Fe_2As_2)(Ca_4(Mg_{0.25}Ti_{0.75})_3O_{7.5})$ below 50 K. The in inset shows a resistivity curve in 0 T up to 300 K.

Fig. 1

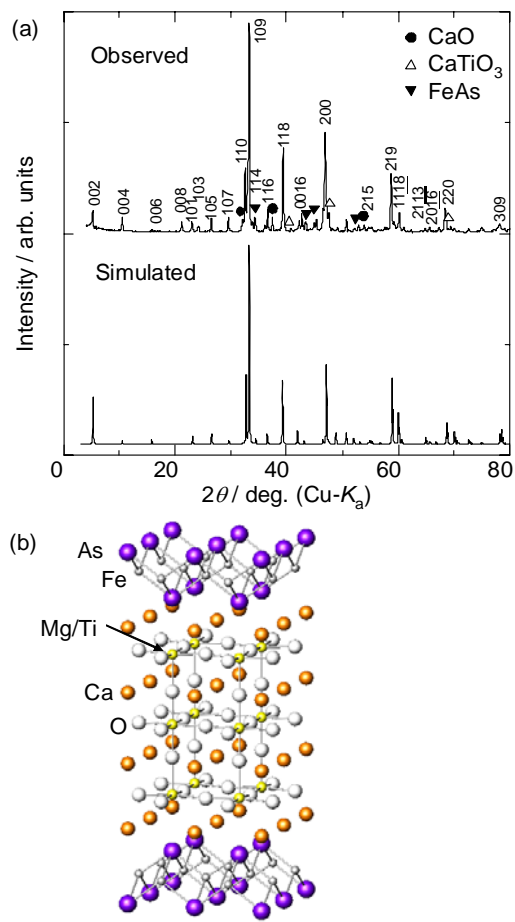

Fig. 2

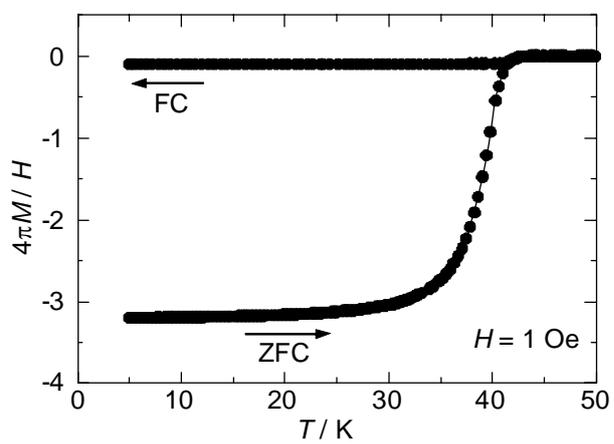

Fig. 3

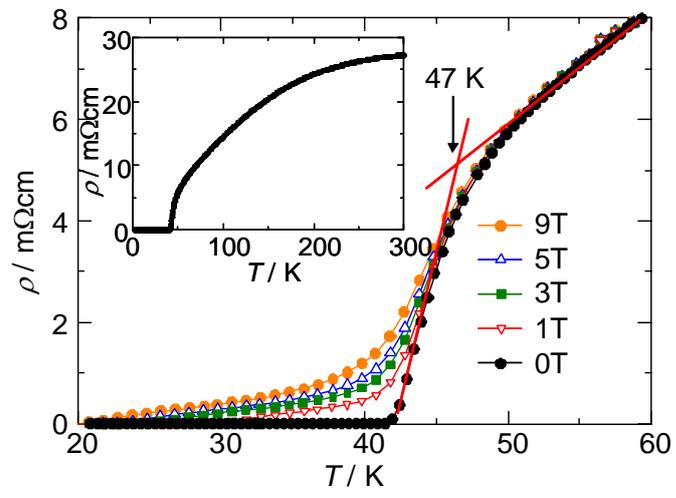